\documentclass[conference]{IEEEtran}
\IEEEoverridecommandlockouts

\usepackage{array}
\usepackage{cite}
\usepackage{amsmath,amssymb,amsfonts}
\usepackage{algorithmic}
\usepackage{graphicx}
\usepackage{textcomp}
\usepackage[dvipsnames]{xcolor}
\usepackage{color,soul}
\usepackage{graphicx, caption, subcaption}
\usepackage{multirow}
\usepackage{multicol}
\usepackage{array}
\newcolumntype{C}[1]{>{\centering\arraybackslash}p{#1}}
\def\BibTeX{{\rm B\kern-.05em{\sc i\kern-.025em b}\kern-.08em
    T\kern-.1667em\lower.7ex\hbox{E}\kern-.125emX}}
\begin{document}

\title{Time Reversal for Near-Field Communications on Multi-chip Wireless Networks
\thanks{\textsuperscript{*}F. Rodríguez-Galán and A. Bandara contributed equally to this work. 
\newline F. Rodríguez-Galán, A. Bandara, E. Alarcón and S. Abadal are with Universitat Polit\`{e}cnica de Catalunya, Barcelona, Spain. E. Pereira de Santana and P. Haring Bolívar are with University of Siegen, Siegen, Germany.
\newline Authors acknowledge support from the European Research Council (ERC) under grant agreement 101042080 (WINC) as well as the European Innovation Council (EIC) under grant agreement No 101099697 (QUADRATURE). Corresponding author: Sergi Abadal (abadal@ac.upc.edu)}
}

\author{F\'atima Rodr\'iguez-Gal\'an*, Ama Bandara*, Elana Pereira de Santana, Peter Haring Bol\'ivar,\\Eduard Alarc\'on and Sergi Abadal}


\maketitle

\begin{abstract}
Wireless Network-on-Chip (WNoC) has been proposed as a low-latency, versatile, and broadcast-capable complement to current interconnects in the quest for satisfying the ever-increasing communications needs of modern computing systems. 
However, to realize the promise of WNoC, multiple wireless links operating at several tens of Gb/s need to be created within a computing package. Unfortunately, the highly integrated and enclosed nature of such computing packages incurs significant Co-Channel Interference (CCI) and Inter-Symbol Interference (ISI), not only preventing the deployment of multiple spatial channels, but also severely limiting the symbol rate of each individual channel. In this work, Time Reversal (TR) is proposed as a means to compensate the channel impairments and enable multiple concurrent high-speed links at the chip scale. We offer evidence, via full-wave simulations at 140 GHz, that TR can increase the symbol rate by an order of magnitude and allow the deployment of multiple concurrent links towards achieving aggregate speeds in excess of 100 Gb/s. Finally, the challenges relative to the realization of TR at the chip scale are analyzed from the implementation, protocol support, and architectural perspectives. 
\end{abstract}


\section*{Introduction} 

The end of Moore's law has forced computer architects to find ways to enhance the performance and efficiency of microprocessors other than by just scaling the transistors. One option is to increase the number of independent processor \emph{cores} in a chip, which leads to the so-called multi-core processors \cite{Nychis2012}; whereas other techniques include the integration of multiple specialized hardware accelerator chips in a single computing package \cite{zimmer20200}. Because of the parallel nature of such architectures and their need to synchronize and share data internally, communications within and across chips become a critical determinant of performance and efficiency in modern computing systems. 



Across academia and industry, most processor families use the concept of Network-on-Chip (NoC) as their interconnect backbone within the chip \cite{Nychis2012}. 
A NoC is a packet-switched network of on-chip routers and links co-integrated with the processor cores. Its use is widespread due to its simplicity and high performance in moderately sized processors. 
However, when scaling towards massive architectures and reaching beyond the boundaries of a simple chip, where wired links become considerably slower \cite{zimmer20200}, NoC exhibits important latency and energy consumption issues. This is especially harmful when communications are collective in nature, since one-to-all and all-to-one patterns (which are common in computing systems) may generate severe congestion inside the NoC.

\begin{figure}[!t]
\centering
\includegraphics[width=1\columnwidth]{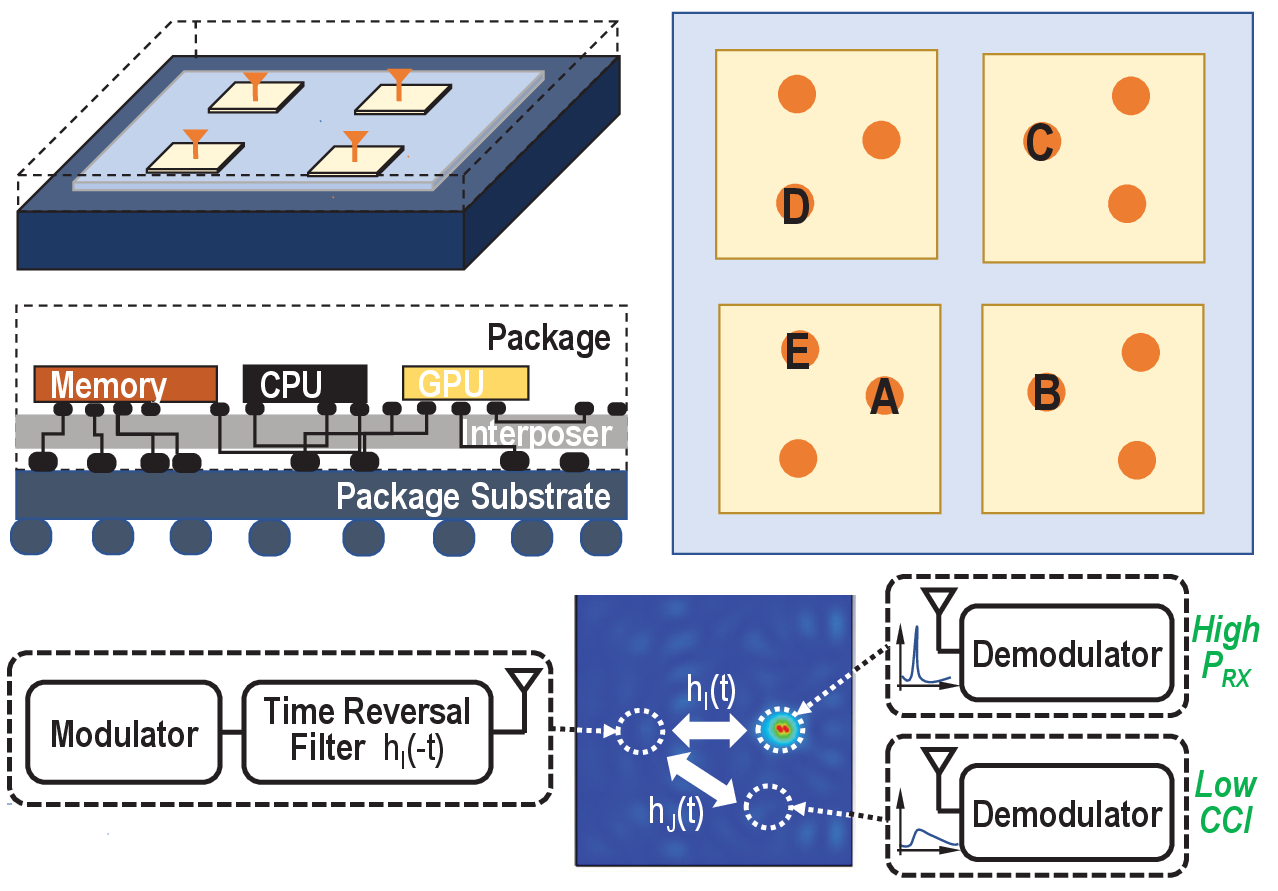}
\vspace{-0.4cm}
\caption{Interposer package with four chiplets and multiple wireless chips-scale links. With time reversal, the transmitted signal is spatiotemporally focused at the intended receiver, achieving high power with a short response. At non-intended receivers, the co-channel interference (CCI) level is diminished.}
\label{fig:trsim}
\vspace{-0.2cm}
\end{figure}

To solve these issues, recent studies have proposed the integration of miniaturized antennas in the millimeter-wave (mmWave) and terahertz (THz) frequency bands to enable Wireless Networks-on-Chip (WNoCs, Figure~\ref{fig:trsim}) \cite{abadal2019wave}. In a WNoC, the data packets are serialized and modulated, and the resulting electromagnetic (EM) waves radiated through the the chip package towards the intended destinations; whereas at the receiving end, the data is demodulated and deserialized. This paradigm is an appealing complement to existing wired solutions because it overcomes several shortcomings of NoCs in the many-core era. For instance, WNoC achieves low latency across the system because only one hop is required to reach any destination, even between chips. Moreover, it provides inherent broadcast capabilities that can be exploited to implement collective communication patterns.


The benefits of WNoC have been explored in multiple works, although most of those studies rely on incorrect channel models. In particular, the availability of multiple wideband channels supporting wireless links at 10 Gb/s is generally assumed, which is incompatible with the highly integrated and enclosed nature of a computing package \cite{timoneda2020engineer}. 
The wireless channel within a computing package is generally extremely reverberating and impaired by near-field effects, leading to long channels and uneven yet widespread energy distributions across the entire space \cite{abadal2019wave}. 
This has two undesired consequences. 
On the one hand, long channel effects will cause severe Inter-Symbol Interference (ISI) as the symbol rate increases and, hence, the symbol duration decreases. This is especially detrimental for the low-order modulations typically proposed in WNoC to avoid power-hungry transceivers, because they require higher symbol rates than high-order modulations to achieve the 10+ Gb/s data rates as assumed in the literature so far. On the other hand, the widespread energy distribution causes Co-Channel Interference (CCI) which prevents the deployment of multiple channels in the same frequency and time and, therefore, severely limits the aggregate data rates achievable in a WNoC. 
In conclusion, solutions are required to mitigate the ISI and CCI in the context of wireless communications within a computing package.


To bridge this gap, this paper presents the vision of Time Reversal (TR)-based wireless communications at the chip scale. TR is a technique wherein the Channel Impulse Response (CIR) is used to compensate the impairments of a given wireless channel by creating a matched filter at the transmitter \cite{lerosey2004time}. To this end, the recorded CIR is reversed in time and transmitted, achieving a spatiotemporal focusing at an intended receiver only, as shown in multiple works at multiple scales and including in the near-field \cite{lerosey2004time, Alexandropoulos2022magazine, Chabalko2016, lerosey2007focusing}. In highly reverberant environments, TR can be used to mitigate the ISI incurred with large delay spreads thanks to its temporal focusing effect. In addition, CCI can also be diminished with TR, given its capacity to focus the EM radiation at the vicinity of the intended receiver only. Hence, TR shows promise in eliminating the current hurdles of the WNoC approach.


Despite its theoretical benefits and recent trials \cite{Alexandropoulos2022magazine}, TR has not been widely adopted in wireless networks. The main reason is that TR needs not only (i) precise and updated Channel State Information (CSI), which is traditionally complex or expensive to obtain in wireless networks, where the channel is constantly changing; but also (ii) a filter capable of adapting to such dynamic channel. In contrast, WNoC appears to be uniquely suited to TR because both the environment and the nodes are static, leading to a channel that is time-invariant and can be pre-characterized. Moreover, the near-field effects naturally arising in WNoC due to the small scale of the chip environment are not an issue, since TR can be applied to channels with near-field effects as long as their consequences are captured in the CIR \cite{Chabalko2016, lerosey2007focusing}. As a result, even subjected to near-field effects, the TR filter can be pre-programmed and re-used, which greatly simplifies the TR approach. 



In summary, in this paper we present and assess the value of TR for wireless communications at the chip scale. We discuss the particularities of the TR technique in this novel wireless scenario, to then assess its capacity of mitigating the ISI and CCI within a multi-chip computing package. We provide full-wave simulations at 140 GHz and show that individual TR links can achieve speeds beyond 10 Gb/s, and that multiple parallel transmissions are possible with low interference, leading to aggregate data rates beyond 100 Gb/s. Finally, we outline some challenges to consider in the design and implementation of TR in future computing systems. This includes considerations of power allocation, compatibility with high-order modulations, dynamic management of multiple parallel TR channels, or the practical implementation and calibration of TR filters.

\begin{table*}[t!]
\begin{center}
\caption{Summary of existing techniques to mitigate ISI and CCI}
\label{tab:table1}\vspace{-0.1cm}
\begin{tabular}{ C{4cm} | C{1.2cm}| C{1.2cm} | C{1.3cm} | C{2cm} |C{2.5cm} | C{2.5cm} 
}
\hline
\multirow{2}{4em}{\bf{Technique}} & \multirow{2}{4em}{\bf{ISI Mitigation}} & \multirow{2}{4em}{\bf{CCI Mitigation}}& \multirow{2}{5em}{\bf{Proposed for WNoC}}&\multirow{2}{8em}{\bf{Reported Band for WNoC}} & \multicolumn{2}{c}{\bf{Remarks Based on WNoC Scenario}}  \\ \cline{6-7} &&&&&\bf{Pros}&\bf{Cons}\\
\hline
 Delay Alignment Modulation (DAM) with Beamforming&\textcolor{ForestGreen}{High}  & \textcolor{ForestGreen}{High}& \textcolor{RedOrange}{No}& N/A & Pre-characterized channel state information. & Noise Amplification.  \\  
 \hline
 Orthogonal Frequency Division Multiple Access (OFDMA)  & \textcolor{ForestGreen}{High} & \textcolor{ForestGreen}{High} &\textcolor{RedOrange}{No}& N/A & Dynamic bandwidth allocation. & Complex signal processing (FFT). \\
 \hline
   Zero Forced Beamforming (ZFBF)  & \textcolor{RedOrange}{No} & \textcolor{ForestGreen}{High} &\textcolor{RedOrange}{No}& N/A & Directional beams. & Noise amplification. Complex signal processing.  \\
  \hline
  Reconfigurable Intelligent
  Surface (RIS) reverberation mitigation & \textcolor{ForestGreen}{High} & \textcolor{ForestGreen}{High} &\textcolor{violet}{Yes}& mmWave \cite{f2022metasurface} & Single RIS serves all links & RIS integration can be complex. \\
   \hline
  Line Coding (RZ) with adaptive thresholds & \textcolor{Apricot}{Medium} & \textcolor{RedOrange}{No} & \textcolor{violet}{Yes} &Sub-Terahertz \cite{timoneda2020engineer} & Simple technique. & Losses \\
   \hline
Time Reversal (TR)  & \textcolor{ForestGreen}{High} & \textcolor{ForestGreen}{High} &\textcolor{violet}{This work}& Sub-Terahertz \cite{Alexandropoulos2022magazine} & Pre-characterized channel state information. Simultaneous ISI/CCI mitigation.  & TR filter implementation and calibration \\
\hline
  \end{tabular}
  \end{center}\vspace{-0.4cm}
\end{table*}

\section*{Time Reversal for on-Chip Communications} 

\label{sec:tronchip}

\noindent \textbf{The chip environment.}
Computing packages hosting multiple interconnected chips represent a new environment for wireless communications in general and for the use of the TR technique in particular. Generally, a chip is composed of stacked metal layers placed on top of a silicon substrate and surrounded by the package, which typically includes a metallic encasement. There are multiple types of packages; in this work, we consider a multi-chip interposer package that is becoming very popular in recent years \cite{abadal2019wave}.

The cross-section of Figure \ref{fig:trsim} shows, in detail, the interposer package used for our simulations. The package is formed by a large silicon chip called interposer, which is placed on top of the package substrate. The interposer sustains a number of chips that could embody a CPU, GPU, memory modules, and others; chips that communicate with each other and with the rest of the computing system through the interposer metallization layers. The chips are typically connected to the interposer using a flip-chip integration technique, which consists in implementing the contacts on their top metallization layer and flipping the chip to interface it with the interposer connectors.
On its turn, a flip-chip is composed of layers of several materials that play different roles in the functionality of the chip. For example, Aluminum Nitride (AlN) is generally used as heat spreader and silicon is used as substrate and for structural support. For further detail on the chip and package structures, we refer the reader to \cite{abadal2019wave}. 

On-chip antennas are integrated within the chips of the computing package to enable wireless communication within and across the chips of the computing system. Several variants can be considered, including planar dipoles, patches, or vertical monopoles \cite{abadal2019wave}. In any case, their radiation will encounter several obstacles in the path to the receiver in the form of media interfaces between the chips, the package, and the package enclosure. All these elements cause a myriad of reflections and delayed time components that contribute to increasing the length of the channel, which can range from a fraction of a nanosecond to a few tens of nanoseconds depending on the locations of the antennas, the dimensions of the package and losses of the materials. Fortunately, the channel response is time-invariant due to the static nature of the environment, and hence can be pre-characterized.

\vspace{0.1cm}
\noindent \textbf{Time reversal within package.}
To address the impairments of the wireless channels within a computing package, we propose to use TR. In essence, TR is signal processing technique that takes full advantage of scattered multipath components in rich scattering environments. There are two main phases in TR communications process, namely, wireless channel characterization and TR precoding, that we illustrate between two arbitrary nodes $A$ and $B$. 

In conventional wireless scenarios, the channel characterization starts by sending a short pulse (whose duration should be a fraction of the maximum frequency to be considered) from node $A$ through the medium and capturing the response of the channel on a given receiver $B$. This process needs to be repeated as often as the channel changes due to mobility or other time-variant factors. In any case, the received signal is recorded, time-reversed, and used as precoding of the data to be transmitted from $B$ back to $A$. As a result of the TR precoding, the transmitted signals by $B$ converge towards $A$ in a focused manner, both in space and time. In other words, TR is able to reduce the length of the $AB$ channel and practically suppress spatial interference to any other nodes of the network.   
Mathematically, this can be formulated through channel correlation; the convolution of a TR transmission with the intended channel maximizes the auto-correlation term, while the convolution of the same transmission with non-intended channels are non-zero cross-correlation terms \cite{rodriguez2023}. It has been found that a reverberant medium helps the time-reversed wave to converge back to the source more accurately \cite{lerosey2004time}. Hence, TR is uniquely suited for the scenario at hand.


In the context of chip-scale communications, TR would be applied as pictured in Figure \ref{fig:trsim}. The characterization of a wireless link (e.g. from $A$ to $B$) can be performed either through simulations or in-situ measurements. The channel response is then computed and used to implement a specific TR filter that can remain fixed over the lifetime of the computing system. This is because the landscape is static, highly controlled, and known beforehand by the designer. This leads to a quasi-deterministic and time-invariant channel, in contrast to traditional wireless networks, where the use of the TR technique is cumbersome because pilots need to be sent frequently back and forth to measure the changes in the medium and guarantee that the channel response estimates are up to date. Besides being time-invariant, the chip environment is also passive, which allows to apply the reciprocity theorem and apply the TR filters in both directions of the same link (in our example, from $A$ to $B$ and vice versa). 

As shown in Table~\ref{tab:table1}, there are alternatives to TR to mitigate ISI and CCI \cite{timoneda2020engineer, zfbf2023, Alexandropoulos2022magazine, Lu2022, f2022metasurface, niknam2016}. However, some of them, such as OFDM and line coding with adaptive thresholds will require highly complex transceivers in order to comply with the bandwidth and error rate requirements of the wireless chip-scale approach. This is an important issue because intricate hardware designs lead to increments in the area and power consumption that are not affordable in a chip. Moreover, most of these techniques do not address ISI and CCI at the same time, limiting their achievable performance. In this context, TR offers a promising compromise between simplicity and performance.

\vspace{0.1cm}
\noindent \textbf{Evaluation Methodology.}
In the subsequent sections, we implement and assess TR for wireless communications at the chip scale and within a computing package. The evaluation scenario, illustrated in Figure~\ref{fig:trsim}, considers the interposer environment with 4 flipped chips, and each chip being augmented with 3 monopole antennas for a total of 12 terminals. Each chip is 4.25$\times$4.25 mm\textsuperscript{2} in size and the interposer covers an area of 11$\times$11 mm\textsuperscript{2}. We assume the use of vertical monopole antennas due to their omnidirectional radiation and increased coupling on the horizontal plane. The antennas are designed to operate at 140 GHz, which implies a wavelength of 2.14 mm in the vacuum of the package and 0.62 mm inside the chip's silicon layer due to its higher dielectric constant. Due to the highly integrated environment, antennas will often be placed in the near field or transition zone of elements that impact on the channel (e.g. some antennas are 0.82 mm, or 1.32$\lambda_{Si}$, apart from the chip edges). 



With this configuration, we first evaluate how TR helps mitigate ISI in single-link configurations within and across chips, to then assess the CCI mitigation in configurations with multiple links operating at the same time. In all explorations, the channel response is obtained for all links in CST Microwave Studio and then exported to Matlab to simulate the process of modulation, TR precoding, and demodulation. With this, the Bit Error Rate (BER) is obtained at the target receivers and for increasing symbol rates. Unless noted, the modulation is Amplitude Shift Keying (ASK) with amplitude ratio of 0.5 and the receiver is implemented as an energy detector. We refer to \cite{rodriguez2023, bandara2023exploration} for more details on the methodology.

\label{sec:backAnt}

\section*{ISI Mitigation: Single-Link Analysis} 
\label{sec:siso}
\begin{figure}[!t]
\centering
\includegraphics[scale=0.5]{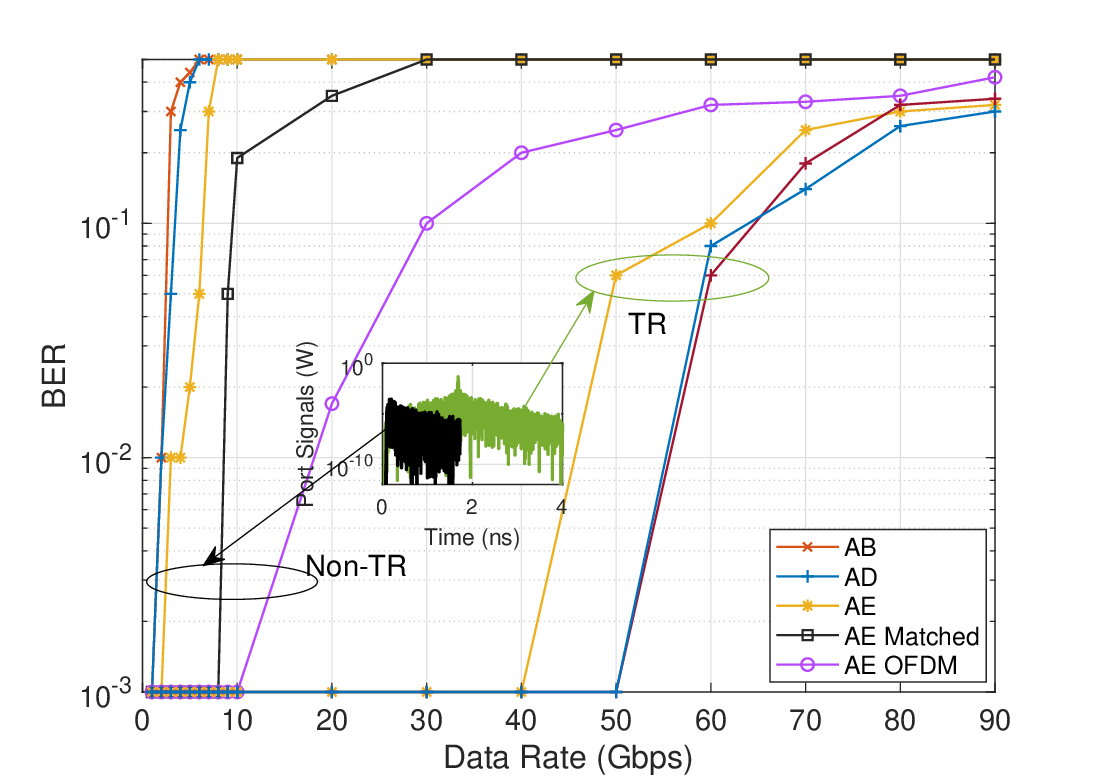}
\vspace{-0.3cm}
\caption{Single-link analysis of time reversal at 140 GHz in an interposer package. Bit error rate is plotted as a function of data rate of ASK modulation with and without time reversal for different links, assuming a transmission power of 10 dBm. For comparison, we also plot OFDM (32 subcarriers) and matched filter applied to one of the links.} The inset shows the energy concentration for a single symbol with and without time reversal.
\label{fig:channel}
\vspace{-0.2cm}
\end{figure}

To quantify the capacity of TR to suppress the ISI, we first evaluate a single-link case. By considering the static and passive nature of the wireless channel inside the chip package, we assume channel reciprocity. For illustration purposes, let us consider antenna $A$ of Figure~\ref{fig:trsim} wants to communicate through the interposer package with antenna $B$ located in an adjacent chip. We obtain the CIR of the link $A\rightarrow B$ with a full-wave simulation. The response is time reversed and convoluted with the modulated data stream $x_{A}$, which will be radiated towards antenna $B$. By the effect of TR, the EM energy will be focused in both space and time around antenna $B$ and $x_{A}$ will be demodulated and retrieved at the receiver. Then, the same process is followed in separate experiments to obtain the response of the $A\rightarrow E$ and $A\rightarrow D$ links to examine the impact of TR on the communication within a single chip and across different chips, respectively. The results are summarized in Figure~\ref{fig:channel}. 

We first compare the channel response before and after applying the TR precoding for a single pulse. It can be observed in the inset of Figure~\ref{fig:channel} how the TR version of the channel response shows a clear peak highly concentrated around 1.8 ns, which is more than two orders of magnitude higher than the peak shown for the non-TR case. The peak is also clearly more concentrated in time, as the delay spread differs by approximately an order of magnitude. 


Second, we compare the impact of TR in a stream of ASK-modulated symbols at different symbol rates, for different intra-chip and inter-chip links. The results, summarized in the main plot of Figure~\ref{fig:channel}, first show how non-TR links cannot exceed speeds of a few Gb/s without quickly degrading the error rate. This is consistent with the delay spread observed for the different links, which is between a fraction of a nanosecond and one nanosecond. We also observe how using TR clearly opens a wide gap and allows to operate at several tens of Gb/s without degrading the link quality. Despite nodes $A$ and $E$ being closer in space, the performance of the intra-chip $A\rightarrow E$ link is slightly worse than the inter-chip counterparts, possibly due to the lossy silicon increasing the path loss and reducing the reverberation, hence reducing the potential of TR in that case. For the sake of comparison, Figure ~\ref{fig:channel} also plots the performance of OFDM with 32 subcarriers and a matched filter applied to one of the links. It is observed how these techniques, which aim at mitigating ISI, are unable to achieve the same performance as TR, that increases the received energy besides reducing the ISI. 

\section*{CCI Mitigation:\\Towards Multiple-Input-Multiple-Output with TR} 
\label{sec:mimo}
\begin{figure*}[!t]
\centering
\begin{subfigure}[t]{0.4\textwidth} 
\includegraphics[width=1\textwidth]{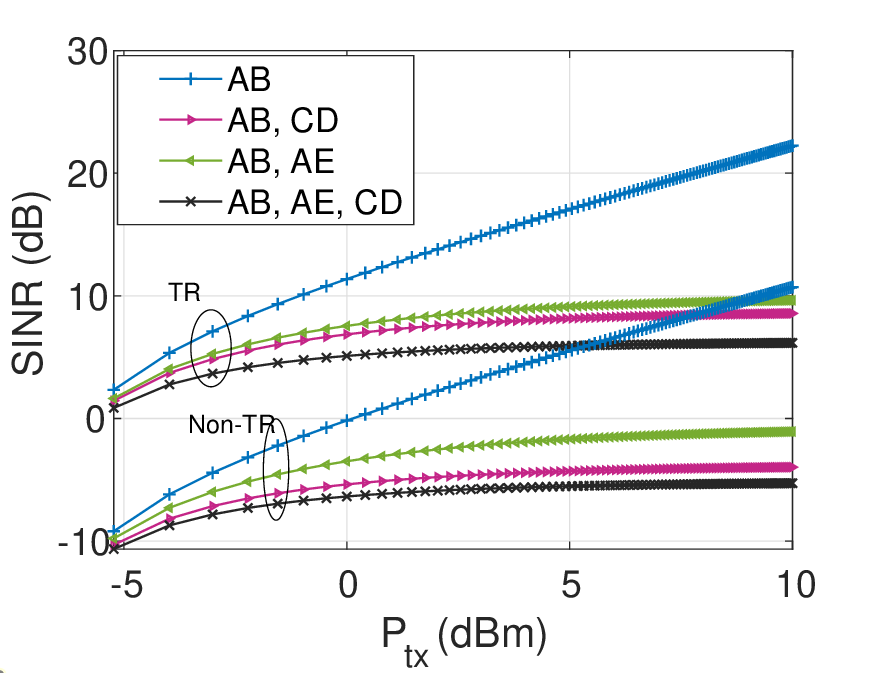}
\caption{\label{fig:multi1}}
\end{subfigure}
\begin{subfigure}[t]{0.4\textwidth} 
\includegraphics[width=\textwidth]{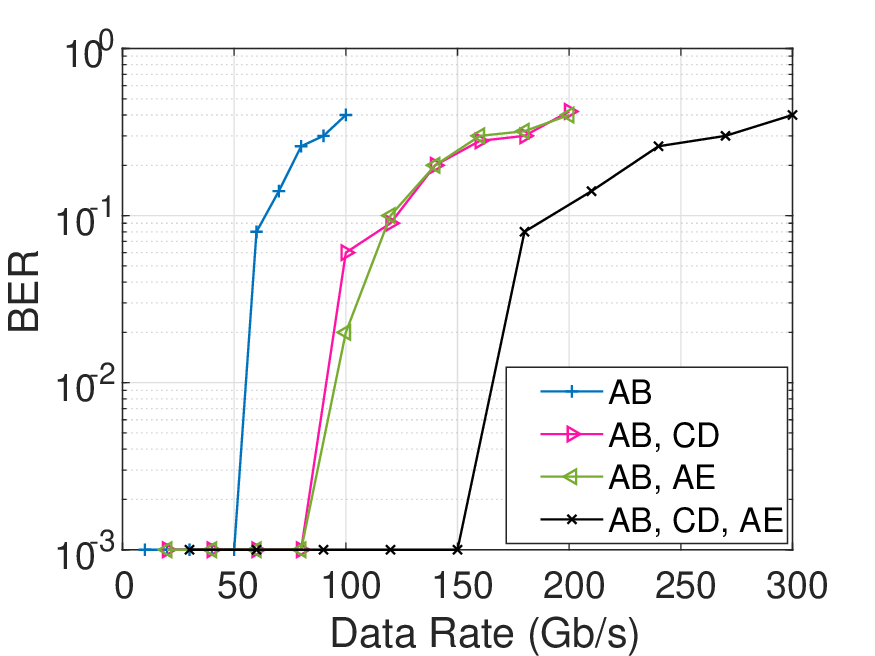}
\caption{\label{fig:multil}}
\end{subfigure}
\vspace{-0.1cm}
\caption{Multi-link analysis of time reversal at 140 GHz considering nodes $A$ and $C$ transmitting concurrently, node $A$ sending different streams to receivers $B$ and $E$ simultaneously, and both at the same time. (a) Signal-to-Interference-Plus-Noise Ratio (SINR) of the multiple configurations. (b) Bit error rate as a function of aggregated data rate of ASK streams assuming a transmission power of 10 dBm per node.} 
\label{fig:multilink}
\vspace{-0.2cm}
\end{figure*}

Once the ISI is mitigated, we focus on the problem of CCI in the case of multiple links being deployed concurrently in time and frequency. Here, let us consider two different cases involving the nodes annotated in Figure~\ref{fig:trsim}. On the one hand, antenna $A$ wants to communicate the signal $x_A$ to antenna $B$, whereas antenna $C$ wants to communicate the signal $x_C$ to antenna $D$. On the other hand, let us consider that $A$ would like to send two different messages, $x_{AB}$ and $x_{AE}$ to nodes $B$ and $E$ respectively.
Hence, we have:
\begin{enumerate}
    \item[(i)] Multiple transmitters sending a single stream to a single receiver each (i.e., $A$ to $B$ and $C$ to $D$).
    \item[(ii)] A single transmitter sending multiple superimposed streams, each with its own TR filter, to multiple receivers simultaneously (i.e., $A$ to $B$ and $E$).
    \item[(iii)] Both (i) and (ii) concurrently.
\end{enumerate}
These cases are very relevant in the on-chip communications, especially the second case, which represents a collective communication pattern generally referred to as \emph{scatter} \cite{rodriguez2023}. Results are summarized in Figure~\ref{fig:multilink}.

In all cases, the simultaneous transmissions incur into different types of CCI as formulated in \cite{rodriguez2023}. However, due to the specific CIR used in TR precoding of the respective channels, the receivers are capable of distinguishing the messages intended for them. Complete suppression of CCI is not possible though, as the different channels involved in the collective communication can show varying degrees of cross-correlation. As we see next, this depends on the relative position of transmitters and receivers, as well as the number of concurrent links. 

The Signal-to-Interference-Plus-Noise Ratio (SINR) of multiple simultaneous parallel communications is measured and illustrated in Figure~\ref{fig:multi1} as a function of transmitted power and compared with the case of no interference and non-TR. First, it is worth noting that TR improves the SINR by around 10 dB in all cases, which will have a remarkable impact on the error rate. Further, we observe that TR cannot completely eliminate the CCI and stalls the SINR as the transmitted power increases. Additionally, when the amount of concurrent transmissions are increased, the SINR is reduced due to the increment of the CCI across channels. However, we do not observe a sizable difference between the SINR of the cases (i) and (ii) described above. 


As TR precoding allows to create reasonably orthogonal spatial channels, we can increase the aggregate bandwidth by increasing the number of channels. As a testament of this, Figure~\ref{fig:multi1} shows the error rate as a function of the aggregated data rate of all the concurrent links for the case of 10 dBm per node. We observe how, indeed, the bandwidth can be pushed beyond 100 Gb/s by adding up to three concurrent channels, combining the two schemes described above. As expected from the SINR analysis, the addition of channels has a certain penalty or overhead; while the single-channel case achieves virtually error-free operation at 50 Gb/s, the dual- and triple-channel cases start saturating at 80 Gb/s and 120 Gb/s, respectively.

\section*{Challenges} 
\label{sec:challenges}





\begin{figure}[!t]
\centering
\includegraphics[width=\columnwidth]{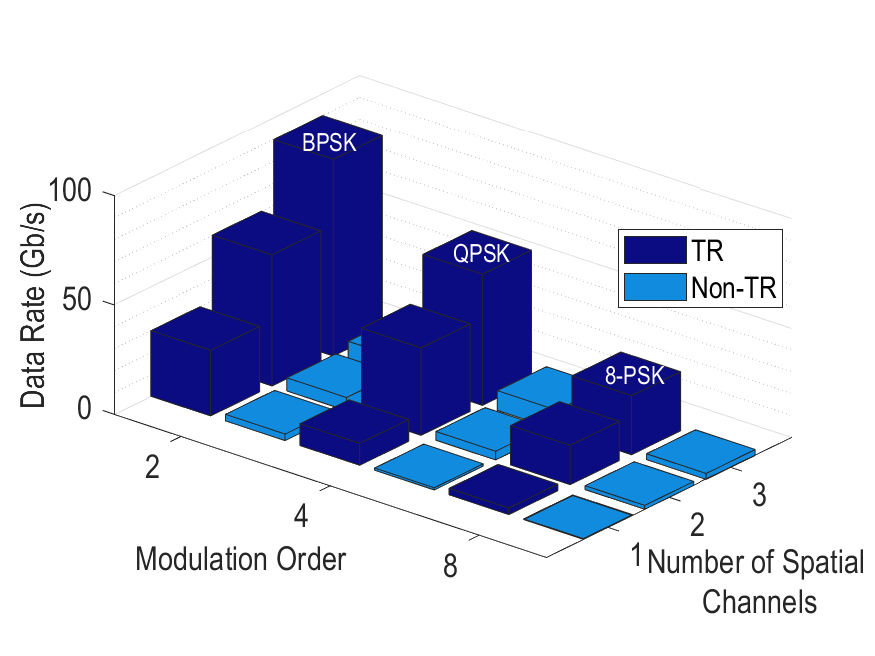}
\caption{Maximum aggregated data rates achieved with BER$<$10\textsuperscript{-3} for phase-modulated streams of increasing modulation order, with and without time reversal, assuming a transmission power of 10 dBm per node.}
\label{fig:datarates}
\end{figure}

\vspace{0.1cm}
\noindent \textbf{Ultra-high Speeds and Ultra-low BER.} In prior sections we have demonstrated how TR can achieve order-of-magnitude improvements in the aggregate data rate of wireless networks at the chip scale. However, the on-chip scenario may require pushing further in the speed of the wireless links. To this end, one may explore the use of high-order modulations to avoid increasing the symbol rate indefinitely. However, as observed in Figure~\ref{fig:datarates}, the use of high-order modulations does not have the expected effect\footnote{While Figure~\ref{fig:datarates} shows the results for phase-modulated streams, a similar tendency can be observed with amplitude modulations.}. This may be caused by the higher SNR requirements of such modulations, which cannot be always compensated with a higher transmitted power due to the CCI, or due to phase distortions. Hence, the challenge is to find techniques to minimize the CCI even further, not only to push the speed, but also to fulfil another of the stringent requirements of on-chip networks: the need to guarantee an error rate comparable to that of wires, far below the $10^{-3}$ level demonstrated in this paper.


\vspace{0.1cm}
\noindent \textbf{Power Allocation.} 
In this paper, we consider that all nodes transmit with the same power and speed. However, power allocation will play a crucial role in collective communications with TR, as they largely determine the SINR at each receiver. The challenge here is to assess, based on the required links at any given time, which is the optimal power allocation per node that will minimize the CCI while guaranteeing an acceptable received power. Fortunately, one can obtain a first-order approximation of the expected CCI by evaluation the correlation between the different channels.
As the environment can be pre-characterized, the channel correlation matrix of a WNoC can be computed once and then incorporated in look-up tables to greatly simplify the power allocation algorithm. This will be crucial in pushing the boundaries of the technique in terms of achievable speed and error rates, as mentioned above.  

\vspace{0.1cm}
\noindent \textbf{Managing Multiple Channels Dynamically.} As we have observed, the cross-correlation between channels will limit the amount of TR links that can be used concurrently. Since the on-chip communication traffic is typically intense and spatio-temporally variable, there is a need for Medium Access Control (MAC) protocols that determine which links to deploy at any given point in time, so that the aggregate data rate is maximized while respecting the bandwidth requirements of the traffic. This process should be performed at runtime. 
Nevertheless, managing multiple concurrent wireless dynamically at runtime is challenging. Due to extreme latency sensitivity of the on-chip communications context \cite{Nychis2012}, existing solutions from conventional 5G/6G networks are inadequate here as they typically involve scheduling or long contention mechanisms. Instead, there have been recent attempts to augment simple and low-latency protocols such as token passing and random access with support for multiple concurrent channels \cite{Ollé2023}. However, such effort does not take into consideration the CCI constraints and hence would lead to reduced performance in a TR-based network. With the benefit of having static and controlled landscape, simple MAC protocols can be designed based on the SINR that can be expected on each parallel receiver. A possible protocol could use a low-speed broadcast channel to announce the communication intent and high-speed TR channels for data communication. In particular, the SINR can be obtained for any combination of nodes that intend to transmit, as well as the correlation index of the channels involved. This information can therefore drive a simple optimization algorithm that chooses the nodes that will use the TR channel, so that the aggregate data rate is maximized at runtime.

\vspace{0.1cm}
\noindent \textbf{TR Filter Implementation.} 
The analysis contained in this work assumes the implementation of a TR filter that recreates the channel response perfectly. 
However, such filter might be considered unfeasible in the on-chip environment, where high temporal resolution is required to capture the channel response effectively, a requisite that conflicts with the evident resource constraints of the on-chip scenario. As per our initial observations, the resolution of the filter in terms of sampling frequency has to be at least commensurate to the symbol rate to be used in the TR link. Below that, the performance degrades significantly due to the appearance of spatial and temporal alias.
Hence, the challenge here resides in the design of practical TR filters that can achieve the spatiotemporal focusing (and the associated ISI and CCI mitigation) with moderate area and power overheads. A gradual reduction of the temporal resolution or the taps of the TR filter may not lead to a graceful degradation of the TR performance, and therefore the tradeoff between filter complexity and TR performance needs to be  studied further in this scenario.


\vspace{0.1cm}
\noindent \textbf{Filter Calibration.} Alongside with TR filter implementation, filter calibration is considered as the foremost step to correct any mismatches between the expected channel response, obtained through simulations, and the actual one. Indeed, the final manufactured and packaged chips may deviate from the nominal or simulated conditions, and consequently the channel may be different from that seen during the design phase. Hence, there should be a protocol to measure the channel once the chip is manufactured and packaged and to adjust the filter accordingly. Fortunately, this is a common practice for the test of chips. In particular, a channel compensation loop could be implemented as a feedback mechanism where initial design conditions could be altered based on the calibration measurements within the integrated package --measurements that can be repeated periodically to guarantee the stability of the TR links.


\section*{Conclusion} 
\label{sec:conclusion}

In this paper, we have proposed the use of TR to mitigate the ISI and CCI effects that severely impair the potential of wireless communications at the chip scale. After a qualitative comparison with other techniques, we assessed the performance of TR using full-wave simulations at 140 GHz in a multi-chip interposer package. The results show that TR thrives with the multipath-rich nature of computer chips, yielding excellent spatiotemporal focusing. We observed improvements of the achievable data rate of over an order of magnitude with respect to non-TR transmissions, and aggregate speeds beyond 100 Gb/s thanks to the deployment of three parallel channels with reduced CCI. Finally, we identified the dynamic management of the multiple TR channels and the practical implementation of the respective precoding filters as two of the most salient challenges of the proposed technique, whose successful implementation could remove the current hurdles preventing the widespread adoption of wireless communications within computing packages.


\bibliographystyle{IEEEtran}
\bibliography{IEEEabrv,bib3}

\section*{} 
\label{sec:bios}
\label{sec:bios}
\begin{IEEEbiographynophoto}{F\'{a}tima Rodr\'{i}guez-Gal\'{a}n} 
is a PhD student at Universitat Polit\`{e}cnica de Catalunya (UPC). Her research interests include electromagnetics, antenna design, and wireless channel characterization.
\end{IEEEbiographynophoto}

\begin{IEEEbiographynophoto}{Ama Bandara} 
is a PhD student at Universitat Polit\`{e}cnica de Catalunya (UPC). Her research interests include signal processing, channel characterization, and MAC protocol design.
\end{IEEEbiographynophoto}

\begin{IEEEbiographynophoto}{Elana Pereira de Santana} 
is a PhD student at the Institute of High Frequency and Quantum Electronics at University of Siegen. Her research interests include 2-D materials and graphene-based antennas for THz communications.
\end{IEEEbiographynophoto}


\begin{IEEEbiographynophoto}{Peter Haring Bol\'{i}var} 
is Chair for High Frequency and Quantum Electronics, University of Siegen. His research interests include terahertz technology, high-frequency electronics, nanotechnology, and photonics.
\end{IEEEbiographynophoto}

\begin{IEEEbiographynophoto}{Eduard Alarc\'{o}n} is an associate professor at the Universitat Polit\`{e}cnica de Catalunya (UPC). His research interests include communications at the nano-scale, wireless energy transfer and computer architecture.
\end{IEEEbiographynophoto}

\begin{IEEEbiographynophoto}{Sergi Abadal} 
is a Distinguished Researcher at Universitat Polit\`{e}cnica de Catalunya (UPC). His research interests include graphene antennas, chip-scale wireless communications, and computer architecture.
\end{IEEEbiographynophoto}


\end{document}